\documentclass[aps,prl,twocolumn,superscriptaddress,showpacs,10pt]{revtex4-1}
\usepackage{graphicx}
\usepackage{calc}
\usepackage{bm}
\usepackage{color}

\begin{document}

\title{Andreev reflection at the edge of a two-dimentional semimetal.}

\author{A.~Kononov}
\affiliation{Institute of Solid State Physics RAS, 142432 Chernogolovka, Russia}
\author{S.V.~Egorov}
\affiliation{Institute of Solid State Physics RAS, 142432 Chernogolovka, Russia}
\author{Z. D. Kvon}
\affiliation{Institute of Semiconductor Physics, Novosibirsk 630090, Russia}
\affiliation{Novosibirsk State University, Novosibirsk 630090, Russia}
\author{N. N. Mikhailov}
\affiliation{Institute of Semiconductor Physics, Novosibirsk 630090, Russia}
\author{S. A. Dvoretsky}
\affiliation{Institute of Semiconductor Physics, Novosibirsk 630090, Russia}
\author{E.V.~Deviatov}
\affiliation{Institute of Solid State Physics RAS, 142432 Chernogolovka, Russia}

\date{\today}

\begin{abstract}
We  investigate  electron transport through the interface between a niobium superconductor and the edge of a two-dimensional semimetal, realized in a 20~nm wide HgTe quantum well. Experimentally, we observe that  typical behavior of a  single Andreev contact is  complicated by both a pronounced zero-bias resistance anomaly and shallow subgap resistance oscillations with $1/n$ periodicity. These results are demonstrated to be independent of the superconducting material and should be regarded as specific to a 2D semimetal in a proximity with a superconductor. We interpret these effects to originate from the Andreev-like correlated process  at the edge of a two-dimensional semimetal.
\end{abstract}


\maketitle

\begin{figure}
\includegraphics[width=\columnwidth]{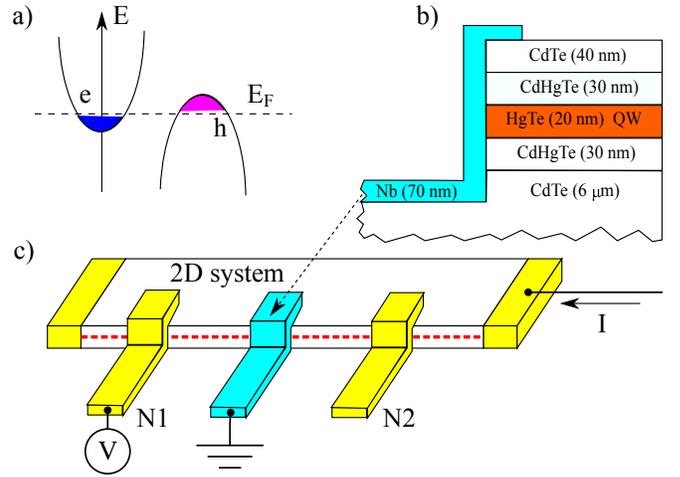}
\caption{(Color online) (a) Schematic energy spectrum of an indirect 2D semimetal. There is a small overlap between the valence and conduction bands, so both electrons (e) and holes (h) contribute to transport. (b) 20 nm $Cd_{0.65}Hg_{0.35}Te/HgTe/Cd_{0.65}Hg_{0.35}Te$ quantum well layer sequence~\protect\cite{growth1,growth2}. The superconducting Nb film  is deposited over the mesa step. (c) Sketch of the sample (not in scale) with electrical connections. The 200~nm deep mesa step is formed by dry Ar plasma etching. The 10~$\mu$m wide superconducting Nb electrode (gray) is placed at the mesa step, with low (2-3~$\mu$m) overlap. Because of the insulating layer on the top of the structure, a {\em side} junction is formed to a quantum well, which is depicted by a red dashed line. Several Au contacts (yellow) are also placed, to obtain normal voltage (N1 and N2) and current  probes.  We study electron transport across one particular SN (Nb -- 2D semimetal) side  junction in a standard three-point technique (see the main text).  
}
\label{sample}
\end{figure}

Recent interest to transport properties of semimetals is connected with a number of new two-dimensional (2D) systems, like bilayer graphene~\cite{graphene1,graphene2}, BiSe thin films~\cite{bise} and wide HgTe quantum wells~\cite{kvon,kvon_s1,kvon_s2}. Similarly to a classical bismuth semimetal, all these materials are characterized by a small overlap between the valence and conduction bands, see Fig.~\ref{sample} (a), so both electrons and holes contribute to transport. In the regime of equal  electron and hole concentrations, while recombination between the carriers from different bands is strongly suppressed, Coulomb correlations become important~\cite{mott,rice}.   

Two-component correlated systems are expected to demonstrate non-trivial physics in  proximity with a superconductor. In the case of a Weyl semimetal~\cite{weyl}, which is an example of the correlated system,  the proximity leads to specular Andreev reflection~\cite{spec} at the interface, or even to superconducting correlations within a semimetal~\cite{corr1,corr2,corr3}. Also, a correlated four-particle Andreev process has been predicted~\cite{bilayer_theor,golubov} at the interface between a superconductor and a bilayer exiton structure~\cite{bilayer_exp}.  One can also expect non-trivial proximity effects for 'classical' 2D semimetals with indirect band structure, because of the allowed exciton formation in the regime of equal electron and hole concentrations~\cite{rice}.

Here, we investigate  electron transport through the interface between a niobium superconductor and the edge of a two-dimensional semimetal, realized in a 20~nm wide HgTe quantum well. Experimentally, we observe that  typical behavior of a  single Andreev contact is  complicated by both a pronounced zero-bias resistance anomaly and shallow subgap resistance oscillations with $1/n$ periodicity. These results are demonstrated to be independent of the superconducting material and should be regarded as specific to a 2D semimetal in a proximity with a superconductor. We interpret these effects to originate from the Andreev-like correlated process  at the edge of a two-dimensional semimetal.


Our $Cd_{0.65}Hg_{0.35}Te/HgTe/Cd_{0.65}Hg_{0.35}Te$ quantum well with (013) surface orientation is grown by molecular beam epitaxy. The layer sequence is shown in Fig.~\ref{sample} (b), a detailed description can be found elsewhere~\cite{growth1,growth2}. At high $d=20.5$~nm width, a 2D system in the quantum well represents an indirect 2D semimetal~\cite{kvon_s1,kvon_s2} with a low overlap between the valence and conduction bands, as depicted in Fig.~\ref{sample} (a). For the undoped well, both electrons and holes contribute to transport. As obtained from standard magnetoresistance measurements, the carriers' concentrations are low enough, about $0.5 \cdot 10^{11}  $cm$^{-2}$ and $1 \cdot 10^{11}  $cm$^{-2}$ for electrons and holes, respectively.  Electrons' mobility is high enough, about $4\cdot 10^{5}  $cm$^{2}$/Vs, because the holes (with lower $5\cdot 10^{4}  $cm$^{2}$/Vs mobility) provide efficient disorder screening~\cite{kvon_scat}.

Fig.~\ref{sample} (c) demonstrates a sample sketch. A 100~$\mu$m wide mesa is formed by 200~nm deep dry Ar plasma etching. We use  magnetron sputtering to deposit a superconducting film  over the mesa step, with low (2-3~$\mu$m) overlap, see Fig.~\ref{sample} (b-c). The 10~$\mu$m wide superconducting electrode is formed by lift-off technique, and the surface is mildly cleaned by Ar plasma before sputtering.  To avoid mobility  degradation, the sample is kept at room temperature during the sputtering process. Ohmic source-drain contacts and the potential probes N1 and N2  are obtained by thermal evaporation of 100~nm thick Au (yellow in Fig.~\ref{sample} (c)). The potential probes are usually placed at a 100~$\mu$m distance from the superconducting electrode. 

Our samples differ by the material of a superconducting contact. It is formed either by a 70~nm thick $Nb$ film, or by a bilayer from a 35~nm thick $Nb$ layer and a 30~nm thick  permalloy $Fe_{20}Ni_{80}$ layer. In both cases the 2D system is in a direct contact with the Nb film, which ensures similar scattering at the SN (superconductor --  semimetal) interface. On the other hand, a premagnetized $Fe_{20}Ni_{80}$  layer partially suppresses superconductivity in Nb, so the bilayer behaves like a Nb superconductor with a strongly reduced gap. 

Without annealing, only a side contact is possible  at the mesa step between the metallic electrode (either superconducting or normal) and a 2D system, because of the insulating CdTe layer on the top of the structure, see Fig.~\ref{sample} (b). 
	We study electron transport across a single SN (Nb  --  semimetal) junction in a standard three-point technique, see Fig.~\ref{sample} (c): the superconducting contact  is grounded; a current is fed through one of the normal Ohmic contacts; the normal contact N1 (or N2) traces the quantum well potential. We sweep a dc current component from -2 to +2~$\mu$A. To obtain $dV/dI(V)$ characteristics,   this dc current is additionally modulated by a low ac (30~pA, 110~Hz) component. We measure both,  dc ($V$) and ac ($\sim dV/dI$), components of the quantum well potential by using a dc voltmeter and a lock-in, respectively. 
	The obtained $dV/dI(V)$ curves are verified to be independent of the mutual positions of the normal Ohmic  contacts, so they only reflect the transport parameters, $V$ and $dV/dI$, of a particular SN (i.e. Nb-2D)  interface. 
 	We check, that the lock-in signal is independent of the modulation frequency in the 60~Hz -- 300~Hz range, which is defined by applied ac filters.  To extract features specific to a 2D semimetal, the measurements are performed at a temperature of 30~mK.


\begin{figure}
\includegraphics[width=\columnwidth]{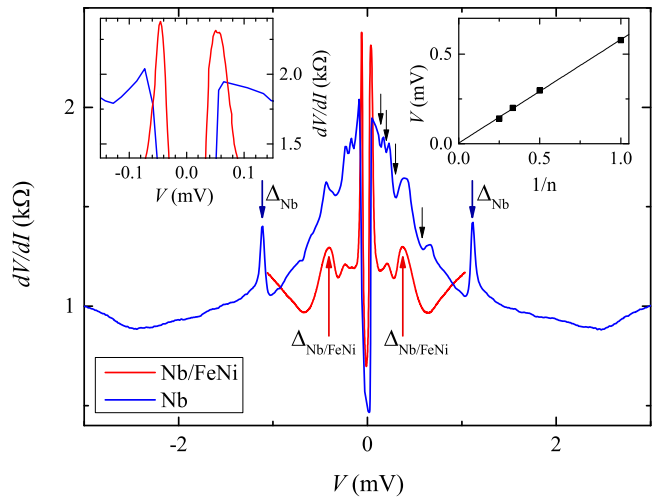}
\caption{(Color online) Examples of  $dV/dI(V)$ characteristics of a single SN junction between the edge of a 2D semimetal and the Nb (blue) or $Nb/FeNi$ (red) superconducting electrodes. Specifics of the 2D semimetal appears (i) in the strong zero-bias resistance anomaly; (ii) in the shallow subgap resistance oscillations (denoted by black arrows). Left inset demonstrates that the anomaly width is almost the same for both superconducting materials. Right inset demonstrates $1/n$ periodicity for the resistance oscillations. All the curves are obtained at   the minimal  temperature $T=30$~mK$<<T_c$ in zero magnetic field.
} 
\label{IV}
\end{figure}

Fig.~\ref{IV} presents the examples of  $dV/dI(V)$ characteristics of a single SN junction between the edge of a 2D semimetal and the  superconducting electrode. The main $dV/dI(V)$ behavior is consistent with the standard one~\cite{tinkham} of a single  Andreev SN junction: every curve demonstrates a clearly defined  superconducting gap (denoted by colored arrows), the subgap resistance  is undoubtedly finite, which is only possible due to Andreev reflection. 
	The superconducting gap for a single-layer Nb junction $\Delta_{Nb}\simeq\pm$~1.15~mV is in a good correspondence with the expected $T_c\approx 9$~K for niobium. The gap is  reduced to $0.3$~meV for a bilayer $Nb/FeNi$ electrode, as obtained from the red arrows in Fig.~\ref{IV}.  
	The  maximum subgap resistance $R_{max}\approx 2$~k$\Omega$  exceeds the normal junction resistance $R_{N}\approx 1$~k$\Omega$, so a single-particle scattering is significant at the Nb-2D interface~\cite{tinkham}. We can be sure, that the upper $Fe_{20}Ni_{80}$ layer is not affecting the interface scattering, because of the similar $R_{max}$ for both superconducting materials. A transmission of the interface $T$ can be estimated~\cite{tinkham} as $R_N/R_{max}\approx 0.5$, which corresponds to the BTK barrier strength~\cite{tinkham} $Z\approx 1$.   

Specifics of the 2D semimetal appears in two striking observations, which can not be expected~\cite{tinkham} for a standard single Andreev contact. (i) The first one is a strong, twice below $R_N$, zero-bias differential resistance anomaly.  The anomaly width and strength are almost the same for a single-layer Nb and for a bilayer $Nb/FeNi$, see Fig.~\ref{IV} and the left inset to it. (ii) The second observation is the shallow subgap resistance oscillations in Fig.~\ref{IV}. They clearly demonstrate $1/n$ periodicity, as depicted in the right inset to Fig.~\ref{IV}. The number of visible oscillations is reduced for $Nb/FeNi$, since the available bias range is effectively diminished by the reduced gap at constant zero-bias anomaly width. These features are gradually diminishing with temperature and disappear at $T=$0.62~K--0.88~K for both $Nb/FeNi$ and $Nb$  junctions, despite the much higher $T_c$ in the last case.

\begin{figure}
\includegraphics[width=\columnwidth]{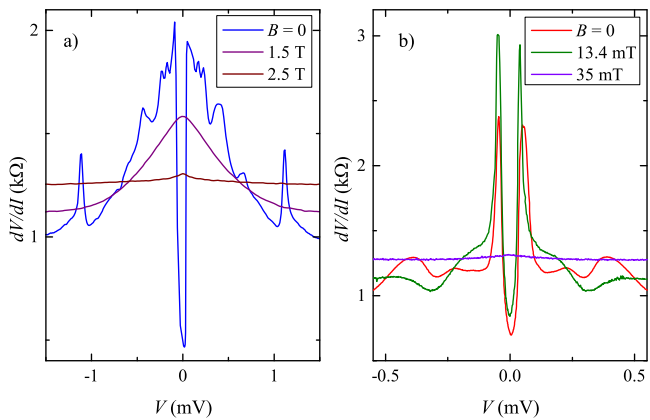}
\caption{(Color online) (a) Effect of the in-plane magnetic field for the SN junction with a single-layer Nb electrode. 
The superconductivity is completely suppressed above 2.5~T, which is consistent with the Nb critical field. In the intermediate fields (1.5~T) the $dV/dI(V)$ curve is non-linear and monotonous. (b)  Because of the  ferromagnetic layer, the superconductivity is fully suppressed by 35~mT field for the SN junction with  a $Nb/FeNi$ bilayer electrode. The zero-bias resistance anomaly, however, can still be seen in the 13~mT field. All the curves are obtained at the minimal $T=30$~mK temperature.
} 
\label{IV_B}
\end{figure}

\begin{figure}
\includegraphics[width=\columnwidth]{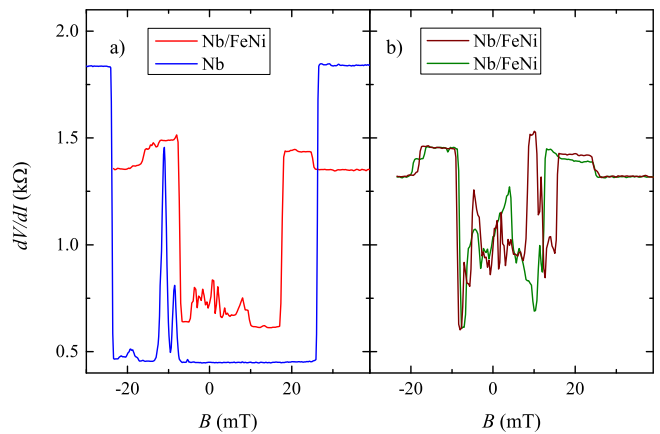}
\caption{(Color online) (a) Threshold suppression of the zero-bias resistance anomaly by the external in-plane magnetic field. The dc bias is fixed at $V=0$ during the field sweep. The resistance is almost constant at low fields, while demonstrates  step-like increase at $\pm 25$~mT for Nb and at (-10,+20)~mT for $Nb/FeNi$. The latter interval is not symmetric, because of the internal magnetization of the permalloy layer. (b) Mesoscopic-like fluctuations within the zero-bias resistance anomaly for a sample with the closely-spaced (5~$\mu$m) potential probe. The fluctuations can not be seen outside the anomaly (in respect to bias and magnetic field intervals), so they are specific for this regime.  All the curves are obtained at the minimal $T=30$~mK temperature.} 
\label{B_switching}
\end{figure}

The effect of the magnetic field is more complicated, see Figs.~\ref{IV_B},~\ref{B_switching}. To avoid orbital effects, the field is oriented within the 2D plane (with $0.5^\circ$ accuracy) along the mesa edge. Thus, it is strictly in-plane oriented also for the superconducting film at the mesa step. We obtain similar low-field results  for the normally oriented magnetic field.

Fig.~\ref{IV_B} (a) demonstrates that the superconductivity can be completely suppressed above 2.5~T for the SN junction with a single-layer Nb electrode, which well corresponds to the Nb critical field (about 3~T for our films). In the intermediate fields, e.g. 1.5~T in Fig.~\ref{IV_B} (a), the $dV/dI(V)$ curve is non-linear and monotonous. The zero-bias anomaly and the oscillations are suppressed simultaneously by very low, below 30~mT, magnetic field.  Qualitatively similar results are obtained for the junction with a $Nb/FeNi$ bilayer, see Fig.~\ref{IV_B} (b).

The low-field behavior of the zero-bias resistance anomaly is shown in Fig.~\ref{B_switching} (a) in detail. We fix the bias $V=0$ and  sweep the magnetic field slowly. For both the superconducting materials, the resistance is almost field-independent  within some interval around zero field.  At the edges of this interval, both $dV/dI$'s demonstrate  step-like increase. This is important, that despite the strongly different $B_c$ for the Nb and $Nb/FeNi$ electrodes, the zero-bias anomaly is characterized by even quantitatively similar behavior in Fig.~\ref{B_switching}. We should connect it with a normal side of the junction, i.e. with a 2D semimetal.

One can see  some irregular $dV/dI(B)$ fluctuations in Fig.~\ref{B_switching} (a) around  zero field. These fluctuations becomes to be extremely strong, if we place the Au potential probe  in a close vicinity (5~$\mu$m) of the superconducting electrode, see Fig.~\ref{B_switching} (b). The fluctuations can not be seen outside the zero-bias anomaly (in respect to bias and magnetic field intervals), so they are specific for this regime.


Thus, for transport through a single SN junction between a superconductor and a 2D semimetal, realized in a wide HgTe quantum well,  two experimental observations have to be understood: (i) the strong zero-bias resistance dip; (ii) the shallow subgap resistance oscillations with the $1/n$ periodicity. These results are independent of the superconducting material and should be regarded as specific to a 2D semimetal in a proximity with a superconductor~\cite{gasb}.

The $dV/dI(V)$ curves in Fig.~\ref{IV} are obtained for a single SN contact. On the other hand, the resonances in the subgap resistance require some space restriction on the normal side of the junction. We can not connect this restriction with trivial disorder: it  can only provide a small, weak antilocalization-like, correction at zero bias, known as disorder-enhanced Andreev reflection~\cite{wal1,wal2}. In contrast, the zero-bias resistance drops twice below the normal junction's value in Figs.~\ref{IV} and \ref{B_switching}.  Moreover, trivial backscattering  can not provide subsequent energy increase in multiple reflections, which is responsible for the $1/n$ periodicity~\cite{tinkham}. Thus, our experiment essentially demands non-trivial (i.e. Andreev-like) scattering on the normal side of the junction, within the 2D semimetal.

\begin{figure}
\includegraphics[width=\columnwidth]{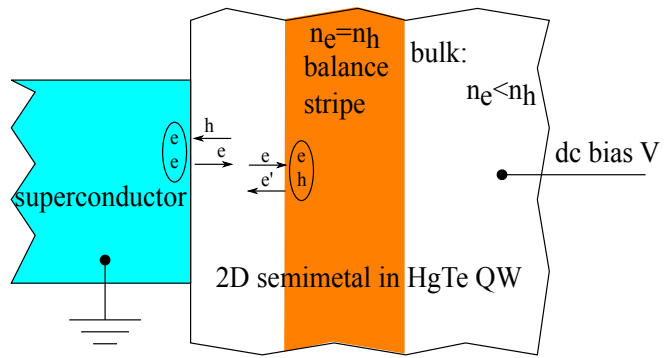}
\caption{(Color online) The edge structure of a 2D semimetal in the vicinity of the superconducting contact. Because of electrostatics~\protect\cite{shklovskii}, the regime of equal concentrations $n_e=n_h$ is stabilized in a stripe of finite width (balance). Andreev-like reflection~\protect\cite{wang,sham} is demonstrated for the balance stripe: to add an electron (e) to the balance regime, an exciton (e-h) should be created, i.e. a hole (h) should also be transferred. It implies reflection of an electron, (e'), which, however, belongs to a different, 'hole',  band of the semimetal spectrum, depicted in Fig.~\protect\ref{sample} (a).
}
\label{discussion}
\end{figure}

Since the data in Fig.~\ref{IV} are qualitatively resemble the typical SNS behavior~\cite{kvon_sns}, we have to connect both experimental findings with scattering on some correlated state near the edge of a 2D semimetal. 
	This correlated state can naturally appear in the regime of equal carriers' concentrations (balance), $n_e=n_h$. The balance regime is necessary realized within the stripe of finite width due to the edge reconstruction~\cite{shklovskii,image02}, as depicted in Fig.~\ref{discussion}. The edge of the sample is a potential barrier for both electrons and holes~\cite{konig}. In our two-component system, the hole concentration is dominant $n_h>n_e$ in the bulk.  The edge potential profile is smooth because of electrostatics~\cite{shklovskii,image02},  so the carriers' concentrations are gradually diminishing to the edge. The dominant (hole) concentration is diminishing faster until the regime of equal concentrations $n_e=n_h$ is reached. This picture agrees with the observed mesoscopic resistance fluctuations in Fig.~\ref{B_switching}: the  balance  stripe is especially sensitive to the long-range potential disorder because of inefficient screening~\cite{halperin} at $n_e=n_h$.

We can propose two possible realizations of a correlated state within the balance $n_e=n_h$ stripe. 

(i) The simplest way is to assume, following Refs.~\onlinecite{corr1,corr2,corr3}, intrinsic superconductive correlations in this $n_e=n_h$ regime. In this case a single SN junction effectively behaves as a SNS-like structure, where Josephson current and multiple Andreev reflections (MAR) are naturally allowed~\cite{tinkham,kvon_sns}. 

(ii) Another candidate for the correlated state is the excitonic phase within the balance $n_e=n_h$ stripe. In this case one can expect both the coherent transport at low energies~\cite{bilayer_theor,golubov} (responsible for zero-bias anomaly) and the specific correlated Andreev-like process at the edge of the stripe. The latter is an analogue of the Andreev process proposed in Refs.~\onlinecite{bilayer_theor} for a bilayer exciton condensate~\cite{bilayer_exp} and of the spinlike Andreev reflection proposed in Refs.~\onlinecite{wang,sham} at the interface of a semimetal and an excitonic phase. 

Until now, there are no experimental confirmations for these predictions (i) and (ii), so both these possibilities should be regarded with care.
	The magnetic field behavior in Fig.~\ref{B_switching} is more consistent with the excitonic assumption (ii).  Indeed, the induced superconductivity (i) is directly connected with the bulk superconductor. On the other hand, the excitonic phase seems to be independent of the superconductor characteristics. In our experiment, the step-like field dependence in Fig.~\ref{B_switching} is almost the same for two strongly different superconducting electrodes $Nb$ and $Nb/FeNi$. Thus, the dependence in Fig.~\ref{B_switching} should be connected with the magnetic field effect on transport to the excitonic phase.

If we consider an electron between the superconductor and the excitonic phase within the balance $n_e=n_h$ stripe, see Fig.~\ref{discussion}, it experience usual Andreev reflection at the superconductor (left in Fig.~\ref{discussion}) interface. At the excitonic (right) interface, charge conservation requires reflection of an electron to add an exciton to the excitonic phase.  This electron, however, belongs to a different, 'hole',  band of the semimetal spectrum~\cite{wang,sham} in Fig.~\ref{sample} (a), since recombination between the carriers from different bands is strongly suppressed in semimetal. This is the key difference from usual backscattering, which makes this reflection  similar~\cite{bilayer_theor,golubov,wang,sham} to usual Andreev process.  In a combination with Andreev reflection at the superconductor interface, the subsequent energy increase is allowed  in multiple reflections, which seems to be responsible for the $1/n$ oscillations periodicity, observed in our experiment.

As a  conclusion, we investigate  electron transport through the interface between a niobium superconductor and the edge of a two-dimensional semimetal, realized in a 20~nm wide HgTe quantum well. Experimentally, we observe that  typical behavior of a  single Andreev contact is  complicated by both a pronounced zero-bias resistance anomaly and shallow subgap resistance oscillations with $1/n$ periodicity. These results are independent of the superconducting material and should be regarded as specific to a 2D semimetal in a proximity with a superconductor. We interpret these effects to originate from the Andreev-like correlated process  at the edge of a two-dimensional semimetal.

We wish to thank  A.M.~Bobkov, I.V.~Bobkova, Ya.~Fominov, V.T.~Dolgopolov, and T.M.~Klapwijk for fruitful discussions.  We gratefully acknowledge financial support by the RFBR (projects No.~13-02-00065 and 13-02-12127), RAS and the Ministry of Education and Science of the Russian Federation under Contract No. 14.B25.31.0007.


\begin{thebibliography}{99}
\bibitem{gasb} We have reproduced these results for a InAs/GaSb semimetal,  so they are also independent of the particular 2D semimetal realization, which will be published elsewhere.
\end{thebibliography}
\end{document}